\documentclass[aps,prl,
twocolumn,
showpacs,amsmath,amssymb,
superscriptaddress,longbibliography]{revtex4-1}
\usepackage{amsmath}
\usepackage{amsfonts}
\usepackage{amssymb}
\usepackage{graphicx}
\usepackage{epsfig}
\usepackage{color}
\usepackage{empheq}
\usepackage{braket}

\newcommand{\RR}{\right}
\newcommand{\LL}{\left}
\newcommand{\m}{\mathrm}
\newcommand{\dg}{\dagger}
\newcommand{\eref}[1]{Eq.~(\ref{#1})}
\newcommand{\fref}[1]{Fig.~\ref{#1}}
\newcommand{\puoli}{\frac{1}{2}}

\begin{document}
\title{Revealing hidden quantum correlations in an electromechanical measurement}

\author{C. F. Ockeloen-Korppi}
\affiliation{Department of Applied Physics, Aalto University, P.O. Box 15100, FI-00076 AALTO, Finland}
\author{E. Damsk\"agg}
\affiliation{Department of Applied Physics, Aalto University, P.O. Box 15100, FI-00076 AALTO, Finland}
\author{G. S. Paraoanu}
\affiliation{Department of Applied Physics, Aalto University, P.O. Box 15100, FI-00076 AALTO, Finland}
\author{F. Massel}
\affiliation{Department of Physics and Nanoscience Center, University of Jyv\"askyl\"a,
P.O. Box 35 (YFL), FI-40014 University of Jyv\"askyl\"a, Finland}
\author{M.~A.~Sillanp\"a\"a}
\email[]{mika.sillanpaa@aalto.fi}
\affiliation{Department of Applied Physics, Aalto University, P.O. Box 15100, FI-00076 AALTO, Finland}
\begin{abstract}

Under a strong quantum  measurement, the motion of an oscillator is disturbed by the measurement back-action, as required by the Heisenberg uncertainty principle. When a mechanical oscillator is continuously monitored via an electromagnetic cavity, as in a cavity optomechanical measurement, the back-action is manifest by the shot noise of incoming photons that becomes imprinted onto the motion of the oscillator. Following the photons leaving the cavity, the correlations appear as squeezing of quantum noise in the emitted field. Here we observe such ``ponderomotive'' squeezing in the microwave domain using an electromechanical device made out of a superconducting resonator and a drumhead mechanical oscillator. Under a strong measurement, the emitted field develops complex-valued quantum correlations, which in general are not completely accessible by standard homodyne measurements. We recover these hidden correlations, using a phase-sensitive measurement scheme employing two local oscillators. The utilization of hidden correlations presents a step forward in the detection of weak forces, as it allows to fully utilize the quantum noise reduction under the conditions of strong force sensitivity.

\end{abstract}
\maketitle

Squeezed states of the propagating electromagnetic field form a fundamental group of non-classical states \cite{Walls1983Squeeze,SirPeter1987}. In a squeezed state, fluctuations of the field in a certain quadrature of the oscillations are 
diminished below the level of vacuum fluctuations. Such property of low noise has raised long-standing interest in precision measurements in optics \cite{Polzik1992Squ,Schnabel2013Qmetro,Bowen2013bio,Bowen2016sqCool,Regal2017} or more recently in the microwave frequency domain \cite{Clerk2015sqQB,Siddiqi2016sqFluo,Esteve2017sqNMR}, since in several detection applications the sensitivity is limited by photon shot noise. This is the situation particularly in the emerging field of gravitational astronomy, where squeezed light will become an indispensable asset in the near future \cite{ligo2013,Schnabel2017}. Furthermore, quantum information processing with continuous variables utilizes multi-mode squeezed states as the essential resource \cite{Polzik1998Teleport,Braunstein2005,Weedbrook}.

On the other hand, squeezed electromagnetic fields are connected to intriguing physics. A squeezed environment can suppress relaxation \cite{Siddiqi2013Squ}, or it can reveal new phenomena \cite{Siddiqi2016sqFluo,Vitali2016sqSense,Teufel2017}. Various nonlinear optical processes such as four-wave mixing or parametric oscillations can produce squeezed light \cite{Squeeze1985,Squeeze1986,Kimble1986Squ,Heidmann87TwoMSq,Kimble1987SqPar}. Following the early work \cite{Yurke1988Sq,Yurke1990Squ}, itinerant microwaves are nowadays routinely squeezed using Josephson parametric amplifiers (JPA) at deep cryogenic temperatures \cite{Lehnert2008Amp,Wallraff2msq,Lehnert2011squ,Deppe2013squ,Huard2012EntUw,Deppe2016SqDispl}.

\begin{figure}[h]
\includegraphics[width=8cm]{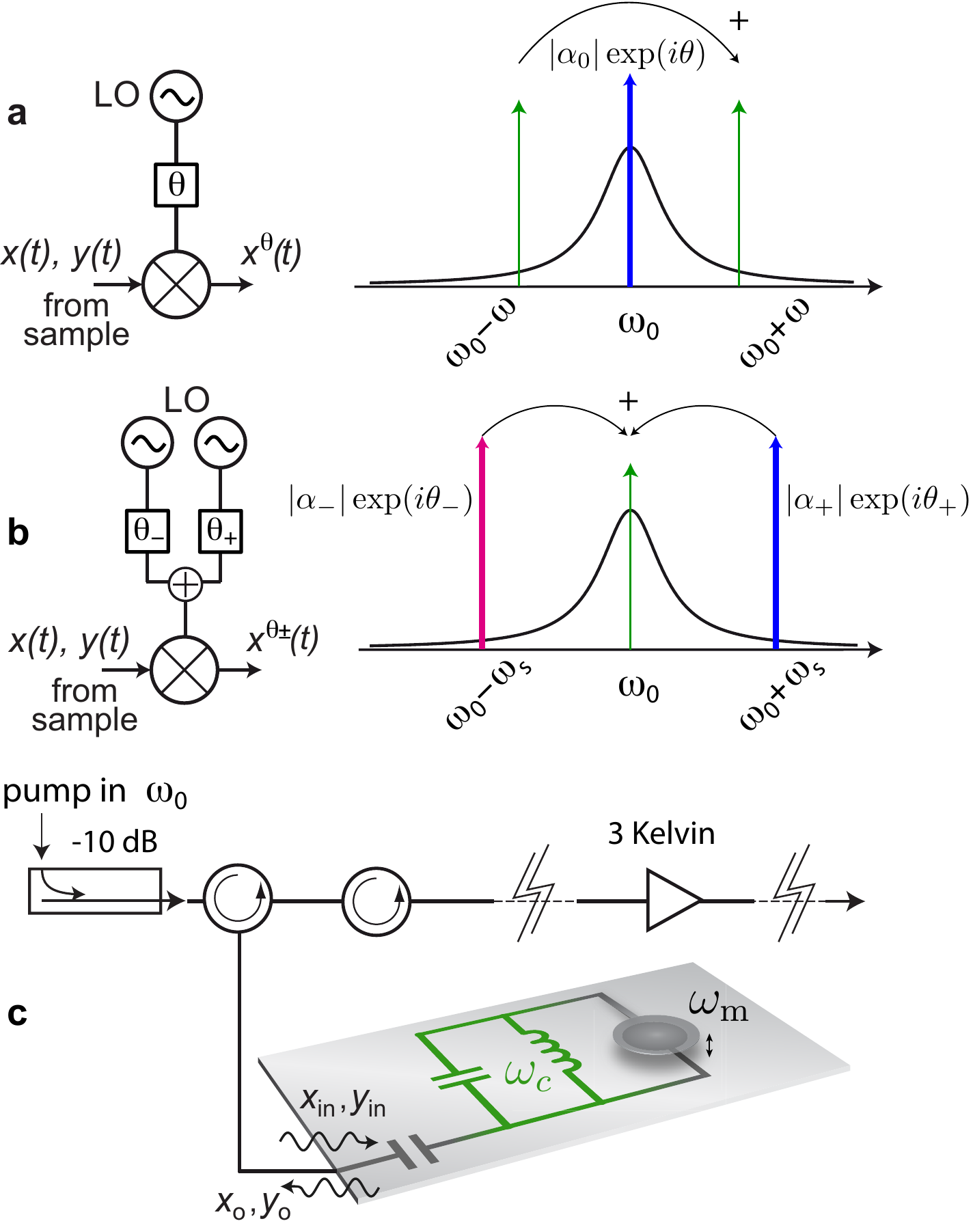}
\caption{\emph{Detection of squeezing correlations.} The signal is injected into a mixer that multiplies it with 
local oscillator (LO) waveforms. \textbf{a}, In the case of standard homodyne detection, the LO is a single sinusoid with frequency $\omega_0$ and phase $\theta$, coinciding with the frequency of the pump used to create correlations in the resonator. This essentially sums up negative and positive frequencies $\pm \omega$ with respect to $\omega_0$.
\textbf{b}, Bi-chromatic LO with two phases can detect complex quantum correlations that may be invisible in homodyne detection. (c) Sketch of the electromechanical experiment, where ponderomotive squeezing of microwaves is created in a sample comprising a micromechanical oscillator  parametrically coupled to a microwave resonator. }
\label{Fig1}
\end{figure}

Cavity optomechanics, which studies the interaction of electromagnetic fields and mechanical oscillations, provides a novel platform to produce squeezed light. The necessary nonlinear mechanism is provided by the radiation-pressure interaction that couples cavity energy to mechanical displacement. Squeezing in optomechanical cavities \cite{Tombesi1994Sq,Heidmann1994sq} arises under an intense measurement that couples 
amplitude fluctuations of an incoming laser to phase fluctuations of the output field. Such ``ponderomotive squeezing'' has recently been produced in several experiments in optics \cite{Atom2012Sq,Painter2013Sq,Regal2013Sq,Schliesser2016,Purdy2017,Kippenberg2017Squ}. In the microwave regime, we mention the realization of strong measurements \cite{Schwab2014QND,Teufel2016ShotN,2BAE} and squeezed microwaves obtained via degenerate parametric amplification \cite{SqueezeAmp}.

In the present work, we show how certain complex-valued quantum correlations, which are hidden from standard homodyne detection, can be recovered by the use of two sinusoidal local oscillators, as recently proposed theoretically by Buchmann \textit{et al.} \cite{Stamper2016syno}. Besides fundamental interest, these hidden correlations are relevant for sensitive measurements since they appear under the condition where the system is the most responsive to forces. Furthermore, as a testbed for this detection setup, we create ponderomotive squeezing in the microwave frequency regime, thereby demonstrating a new approach to create squeezed microwaves, distinct from JPA or from that realized in Ref.~\cite{SqueezeAmp}.

At optical frequencies, squeezing is detected using homodyne detection. Within the theoretical framework introduced by Glauber \cite{Glauber:1963ei}, photodetectors are sensitive to the even normal-ordered correlators of the electromagnetic field. In order to measure the expectation values of a general quadrature operator $X^{\theta}(\omega)=\frac{1}{2}\left(a^{\dagger}e^{i\theta}+a e^{-i\theta} \right)$ and its correlation functions, it is necessary to mix the incoming signal with a local field $b$ (the local oscillator, LO). Depending on the state of the local field, it is possible to access the expectation value of the original quadratures through an intensity measurement of the mixed signal. 
The inclusion of low-pass filters in the measurement process allows us to extend the  applicability of this description to the microwave regime.

In the standard (balanced) homodyne detection setup, the LO is chosen to be in a coherent state at a specific frequency, i.e. $\braket{b}=\alpha_0 \exp\left(-i \omega_0 t\right)$. In this case, it is possible to show that homodyne detection allows access to the following frequency-domain correlator
\begin{equation}
\begin{split}
S_X^\theta (\omega) = & S_X(\omega) \cos^2 \theta + S_Y(\omega) \sin^2 \theta + \\
+ & 2 \mathrm{Re} \LL[ S_{XY} (\omega) \RR] \sin \theta \cos \theta \,.
\end{split}
\label{eq:homo}
\end{equation}
Here, $S_{X}(\omega) =\frac{1}{2} \langle \left\{X(\omega),X(-\omega)\right\} \rangle$, similarly for $S_Y$, and the cross-spectrum is $S_{XY}(\omega) =\frac{1}{2} \langle \left\{X(\omega), Y(-\omega)\right\} \rangle$. Since the cross-spectrum of the two complex-valued frequency-domain quantities is usually not real, information may be lost in homodyne detection. This is represented in \fref{Fig1}a, which shows how positive and negative sideband frequencies sum up.

There has been little earlier discussion on the recovery of complex-valued squeezing correlations hidden to ordinary homodyne detection. Quantum squeezing in the hidden regime has been achieved in optics in one experiment via a modification of the two sidebands \cite{Villar2013syno}.
In the classical limit, an analogous noise reduction was recently obtained in a cavity optomechanical experiment \cite{Marin2018} using digital filtering. 
Quite recently it was proposed \cite{Stamper2016syno} that the complex correlations could be accessed with a bi-chromatic LO, i.e., $\alpha_0(t) = |\alpha_-|  \exp (-i \omega_s t - i\theta_-) + |\alpha_+| \exp (i \omega_s t - i\theta_+)$, as displayed in \fref{Fig1}(b). 
The resulting noise spectrum can be written in a simple form at zero frequency (in the frequency frame oscillating at $\omega_0$) \cite{SM},
\begin{equation}
\begin{split}
S_X^{\theta_\pm}(0) = & |\alpha_X|^2 {\cal C}_{11}(\omega_{s}) + |\alpha_Y|^2 {\cal C}_{22}(\omega_{s}) + \\
& + 2 {\rm Re}[\alpha_X^{*} \alpha_Y^{*}  {\cal C}_{12}(\omega_{s})] \,,
\label{eq:bi}
\end{split}
\end{equation}
where $\alpha_X = \frac{1}{\sqrt{2}}(\alpha_{+} + \alpha_{-}^{*})$, and $\alpha_Y = \frac{i}{\sqrt{2}}(\alpha_{+}^{*} - \alpha_{-})$. We  have  denoted the respective spectra with the correlation matrix elements ${\cal C}_{ij}$, for example ${\cal C}_{12} \equiv  S_{XY} $. 
The relation given by \eref{eq:bi} allows us to interpret $S_X^{\theta_\pm}(0)$ as a quadratic form in the variables $\alpha_X$ and $\alpha_Y$ associated  with the matrix ${\cal C}_\mathrm{ij}(\omega)$. The measurement of $S_X^{\theta_\pm}(0)$ accesses the smallest (largest) eigenvalue of  ${\cal C}_\mathrm{ij}(\omega_s)$  by the choice of $\alpha_X$, $\alpha_Y$ in such a way that the vector $(\alpha_X, \alpha_Y)^\mathrm{T}$  corresponds to the eigenvector associated to the smallest (largest) eigenvalue of
${\cal C}_\mathrm{ij}(\omega_s)$, thereby revealing correlations hidden to homodyne detection.

\begin{figure*}
\includegraphics[width=18cm]{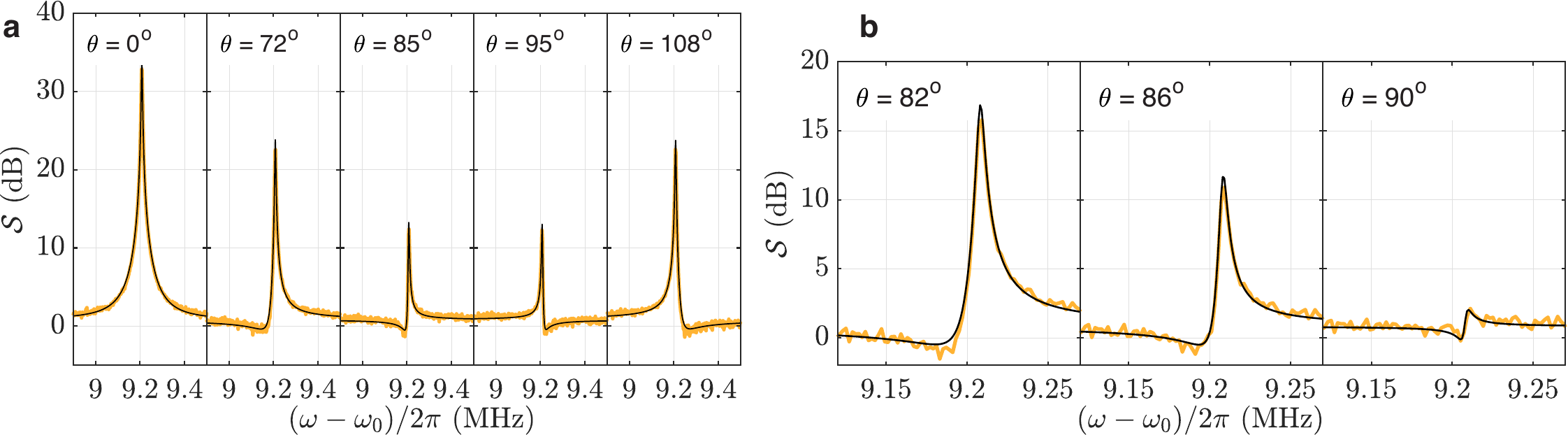}
\caption{\emph{Ponderomotive squeezing of microwaves.} \textbf{a}, Homodyne spectrum, Eqs.~(\ref{sqdef},\ref{sqresult}), as referred to the input of the cryogenic amplifier. The phases of the sinusoidal local oscillator are written in the panels. \textbf{b}, Detailed views of the phases around $\pi/2$ that display the quantum squeezing. The thin black lines are theoretical fits. The experimental parameters are $G/2\pi \simeq 728$ kHz, $n_m^T \simeq 517$, $n_c^T \simeq 0.07$, $\Delta/2\pi \simeq -620$ kHz. }
\label{FigPonder}
\end{figure*}

The realization of the measurement leading to \eref{eq:bi} is not restricted to a specific system. We choose to work in a generic cavity optomechanical setup \cite{Stamper2016syno}, where the observability of the interesting quantities is expected to be well within reach. The interaction between the electromagnetic cavity (frequency $\omega_c$, damping rate $\kappa$, and the  field operators $a^\dg$, $a$), and the mechanical oscillator (frequency $\omega_m$, damping rate $\gamma_m$, operators $b^\dg$, $b$) is of the form $g_0 a^\dg a (b^\dg + b)$, where the single-photon coupling  $g_0 \ll \kappa$ is the small parameter. In order to obtain an effectively strong electromechanical coupling, a strong sinusoidal pump tone at frequency $\omega_0 \simeq \omega_c$ is injected in the system. One can write the pump frequency using the detuning $\Delta \equiv \omega_0 - \omega_c$. The pump induces a photon number $n_c$ in the cavity, and consequently one obtains a linearized interaction $G (a^\dg + a) (b^\dg + b)$ with the effective coupling $G = g_0 \sqrt{n_c} \gg g_0$.

The dynamics is commonly written by the use of input-output theory of optical cavities for the linearized system. There is incoming electromagnetic noise  at the device input,  
which in the present case of low temperature $k_B T \ll \hbar \omega_c$ is composed of vacuum noise having the quadratures $x_\mathrm{in}(t)$ and $y_\mathrm{in}(t)$.
The mechanical oscillator phonon number $n_m^T \simeq k_B T /\hbar \omega_m$, on the other hand, is relatively far from the ground state. The field quadratures leaking out from the cavity receive a contribution by the classical dynamics in the electromechanical system, but also by the fundamental measurement quantum back-action. In order to obtain the output field  $X_{\mathrm{o}}(t)$, $Y_{\mathrm{o}}(t)$, for simplicity we do not in the following write down the mechanical thermal noise. The frequency-domain quadratures are \cite{SM}
\begin{subequations}
  \begin{alignat}{2}
     &X_{\mathrm{o}}(\omega) &&=  \mathcal{A}_{\mathrm{XX}} X_\mathrm{in} (\omega)
+ \mathcal{A}_{\mathrm{XY}} Y_\mathrm{in}(\omega)
 \label{eq:XYout1} \\
     &Y_{\mathrm{o}}(\omega) &&=   \mathcal{A}_{\mathrm{YX}} X_\mathrm{in}(\omega) + \mathcal{A}_{\mathrm{YY}} Y_\mathrm{in}(\omega)
 \label{eq:XYout2} 
  \end{alignat}
\end{subequations}
where the coefficients are
%
\begin{equation}
\begin{split}
   \label{eq:Axy1} 
    &\mathcal{A}_{\mathrm{XX}} =\kappa \eta\chi_\mathrm{c} \left(\frac{\kappa}{2} -i \omega \right)-1, 
   \quad \quad \mathcal{A}_{\mathrm{XY}}  = \kappa \eta\chi_\mathrm{c} \Delta,
   \\
   %
   &\mathcal{A}_{\mathrm{YX}} =-\kappa \eta \chi_\mathrm{c} 
\left( \Delta +4 G^2 \omega_\mathrm{m} \chi_\mathrm{m} \right),\\
   &\mathcal{A}_{\mathrm{YY}} = \kappa \eta \chi_\mathrm{c} \left(\frac{\kappa}{2}-i\omega \right)-1 \,,
  \end{split}
\end{equation}
and $\eta = \left[1+ 4 G
  \omega_\mathrm{m}\chi_\mathrm{m}\Delta\chi_\mathrm{c}\right]^{-1}$, $\chi_\mathrm{c} = \left[\left(\frac{\kappa}{2}-i\omega\right)^2
      +\Delta^2\right]^{-1}$, $\chi_\mathrm{m} = \left[\left(\frac{\gamma_m}{2}-i\omega\right)^2
      +\omega_\mathrm{m}^2\right]^{-1}$.

Ponderomotive squeezing qualitatively arises because the measurement backaction affects each of the output quadratures in a distinct way. The case $\Delta=0$ represents the most direct example, with  $A_{\mathrm{XX}}=A_{\mathrm{YY}}$ and $A_{\mathrm{XY}}=0$, but $A_{\mathrm{YX}}\neq 0$. This, on one hand, implies squeezing; $\braket{X_{\mathrm{o}}(\omega)X_{\mathrm{o}}(-\omega)}\neq \braket{Y_{\mathrm{o}}(\omega)Y_{\mathrm{o}}(-\omega)}$ and, on the other hand, the appearance of nontrivial correlations among quadratures; $\braket{X_{\mathrm{o}}(\omega)Y_{\mathrm{o}}(-\omega)}\neq 0$. The emergence of nontrivial correlations is related to the fact that it is not possible to recast Eqs.~(\ref{eq:XYout1},\ref{eq:XYout2}) in diagonal form through a pair of orthogonal transformations of the input and output quadratures. Another example of a system for which Eqs.~(\ref{eq:XYout1},\ref{eq:XYout2}) cannot be written in diagonal form --hence leading to a mixing of the quadrature signals and a complex-valued ${\cal C}_{12}(\omega)$-- is represented by a phase-mixing amplification (PMA) setup \cite{SqAmpTheory,SqueezeAmp}, showing how PMA and hidden correlations are closely related concepts.

Our experimental scheme is that of microwave cavity-optomechanics, see \fref{Fig1}c. We use a superconducting on-chip cavity resonator (frequency $\omega_c/2\pi \simeq  7.31$ GHz) coupled to a mechanical drum oscillator that has the frequency $\omega_m/2\pi \simeq 9.204$ MHz and the damping rate $\gamma_m/2\pi \simeq 120$ Hz. The single-sided cavity is strongly coupled to the measurement port through the coupling rate $\kappa_E/2\pi \simeq 27.7$ MHz. The cavity also has internal losses at the rate $\kappa_I/2\pi \simeq 100$ kHz, and the cavity losses sum up to $\kappa = \kappa_I + \kappa_E \simeq 2\pi \times 27.8$ MHz. The parameters are selected such that we operate somewhat in the bad-cavity limit $\kappa \gg \omega_m$ and the cavity responds fast to the mechanical fluctuations induced by the incoming shot noise, and hence the amount of squeezing is optimized.

\begin{figure}[h]
\includegraphics[width=8.5cm]{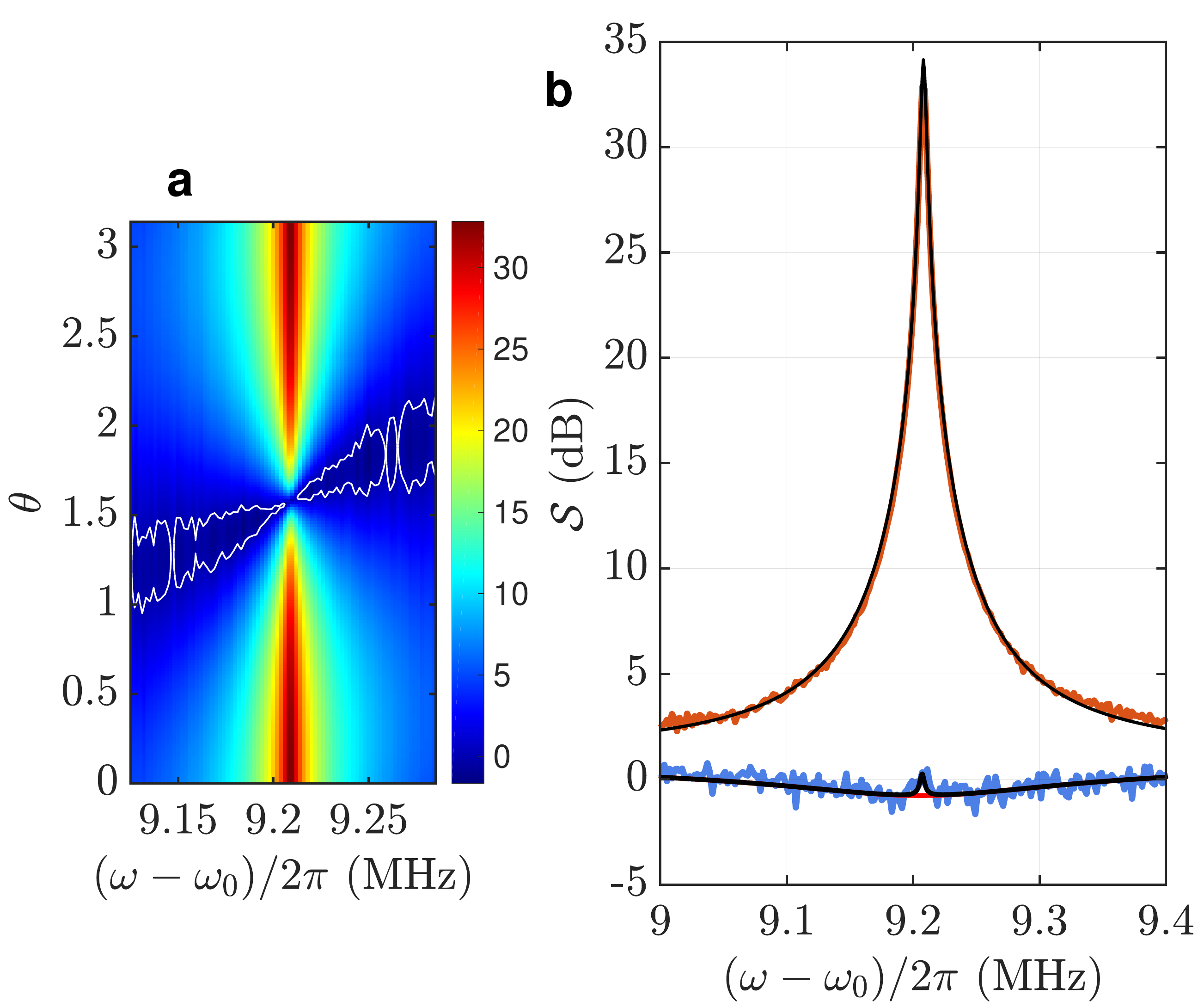}
\caption{\emph{Squeezing eigenvalues.} \textbf{a}, Homodyne noise spectrum shown as a colormap. We observe quantum squeezing (value below 0 dB) inside the white contours. \textbf{b}, Maximum (red) and minimum (blue) envelope of the spectrum with respect to the local oscillator phase $\theta$ from the data in \textbf{a}. The black lines are theory predictions, and the red theory line is the lower eigenvalue of the correlation matrix.
}
\label{FigPonderB}
\end{figure}

The output signal from the sample is directed via isolators and superconducting cables towards the amplifier at 3 Kelvin. To avoid saturating the amplifier, we cancel the strong pump tone by summing up the original signal applied via a $-20$ dB directional coupler, similar to our earlier works  \cite{Squeeze,2BAE,SqueezeAmp,Entanglement}.  
We detect the squeezing in the plane immediately preceding the 3~K amplifier.
This is a regular phase-preserving HEMT amplifier, and adds around $N_{\m{HEMT}}  \approx 10$ quanta of noise to the signal. At  room temperature, the signal is further amplified  and digitized in a signal analyzer that provides the in-phase and out-of-phase quadratures. 

Squeezing can be quantified as the noise in one quadrature $N^\theta$ in units of the vacuum noise in that quadrature, $N_{zp} ^\theta = \frac{1}{4}$,
\begin{equation}
\label{sqdef}
\mathcal{S} = \frac{N^\theta }{N_{zp}^\theta} \, ,
 \end{equation}
therefore $\mathcal{S} < 1$ (or 0 dB) entails quantum squeezing. To infer the squeezing, we measure the quadrature spectral density $S^\theta$ with the pump tone on, and in a separate measurement $S_{\m{off}}^\theta$ with the pump tone off. This allows us to use $N_{\m{HEMT}}$ as a reference, which remains unchanged in both measurements. Hence, \eref{sqdef} becomes
\begin{equation}
\label{sqresult}
\mathcal{S} =  \frac{ N_{\m{HEMT}}^\theta }{N_{zp}^\theta}\LL( \frac{  S^\theta } {S_{\m{off}}^\theta} - 1 \RR),
\end{equation}
where $N_{\m{HEMT}}^\theta = \puoli N_{\m{HEMT}}$ is the HEMT noise in one quadrature.

When using a single LO at the pump frequency (regular homodyne detection, \fref{Fig1}a) we observe a strong phase dependence in the output noise. At LO phase values around $\theta \simeq \pi/2$, we observe a maximum quantum squeezing of $1.1$ dB, see \fref{FigPonder}. The theoretical model shows a good agreement. For the fits, we used as free parameters the mechanical and cavity noise temperatures. The effective coupling and pump detuning are within $\simeq 5$ \% of values calibrated via sideband cooling.

To proceed towards detecting the hidden quantum correlations, we next explore the possible values of the spectrum $\mathcal{S}$ at a given frequency. We make a dense scan of the LO phase (\fref{FigPonderB}a), and record at each frequency the minimum and maximum values of the spectrum. As displayed in \fref{FigPonderB}b, at the mechanical resonance frequency the minimum envelope develops a peak, which does not exhibit squeezing. In the figure we have also plotted the theoretically expected eigenvalues of the correlation matrix, and one can see that the mentioned peak clearly rises higher than the smaller eigenvalue.

\begin{figure}[h]
\includegraphics[width=8cm]{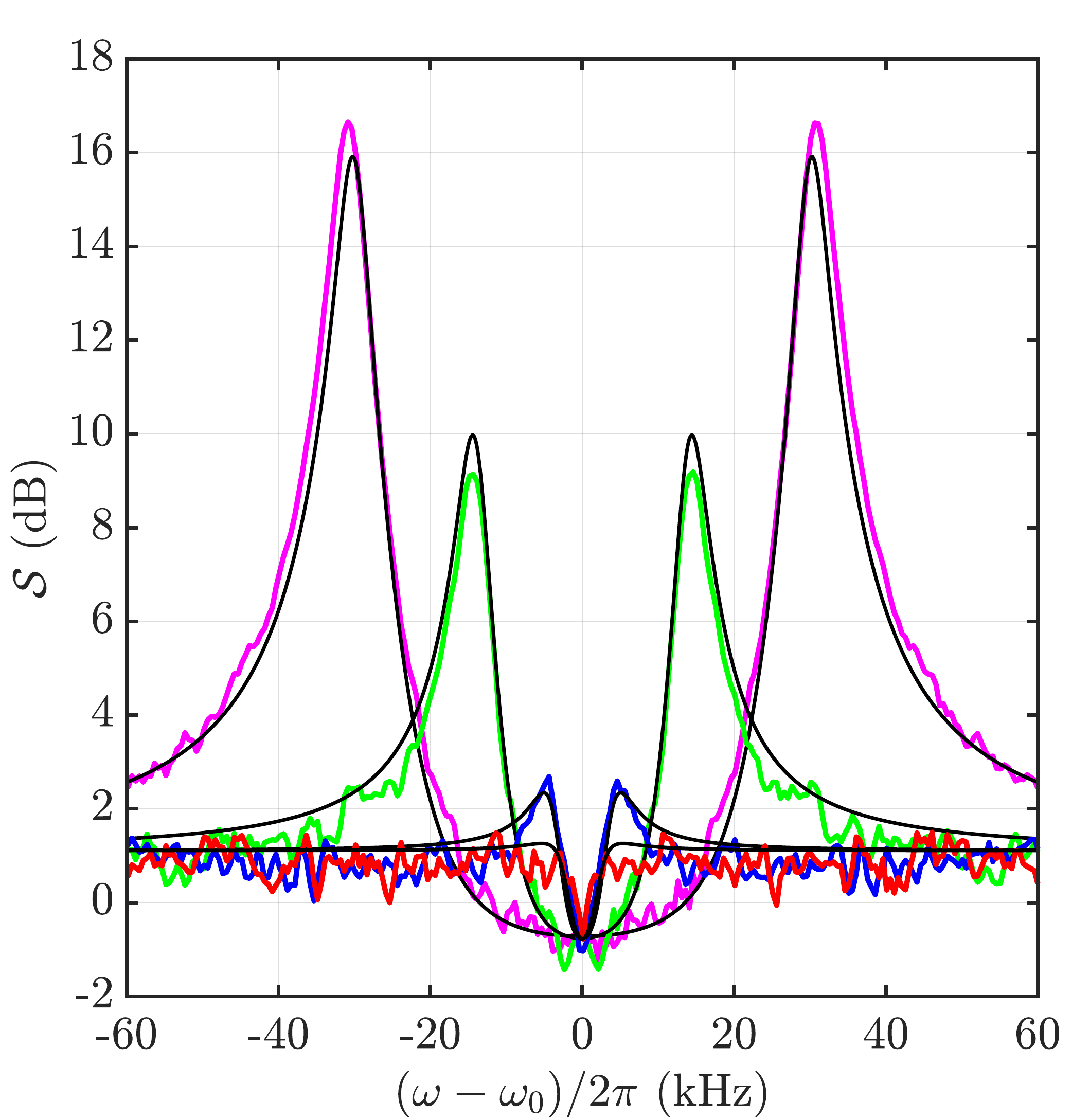}
\caption{\emph{Complex squeezing spectrum.} The detuning of the two LO's are varied between $(\omega_s - \omega_m)/2\pi = [-31, -15, -4.5, 1]$ kHz from top to bottom (magenta, green, blue, red). The phase is optimized in order to minimize the noise for each detuning value. Black lines are theoretical fits.}
\label{FigCompSpec}
\end{figure}

\begin{figure}[h]
\centering
\includegraphics[width=9cm]{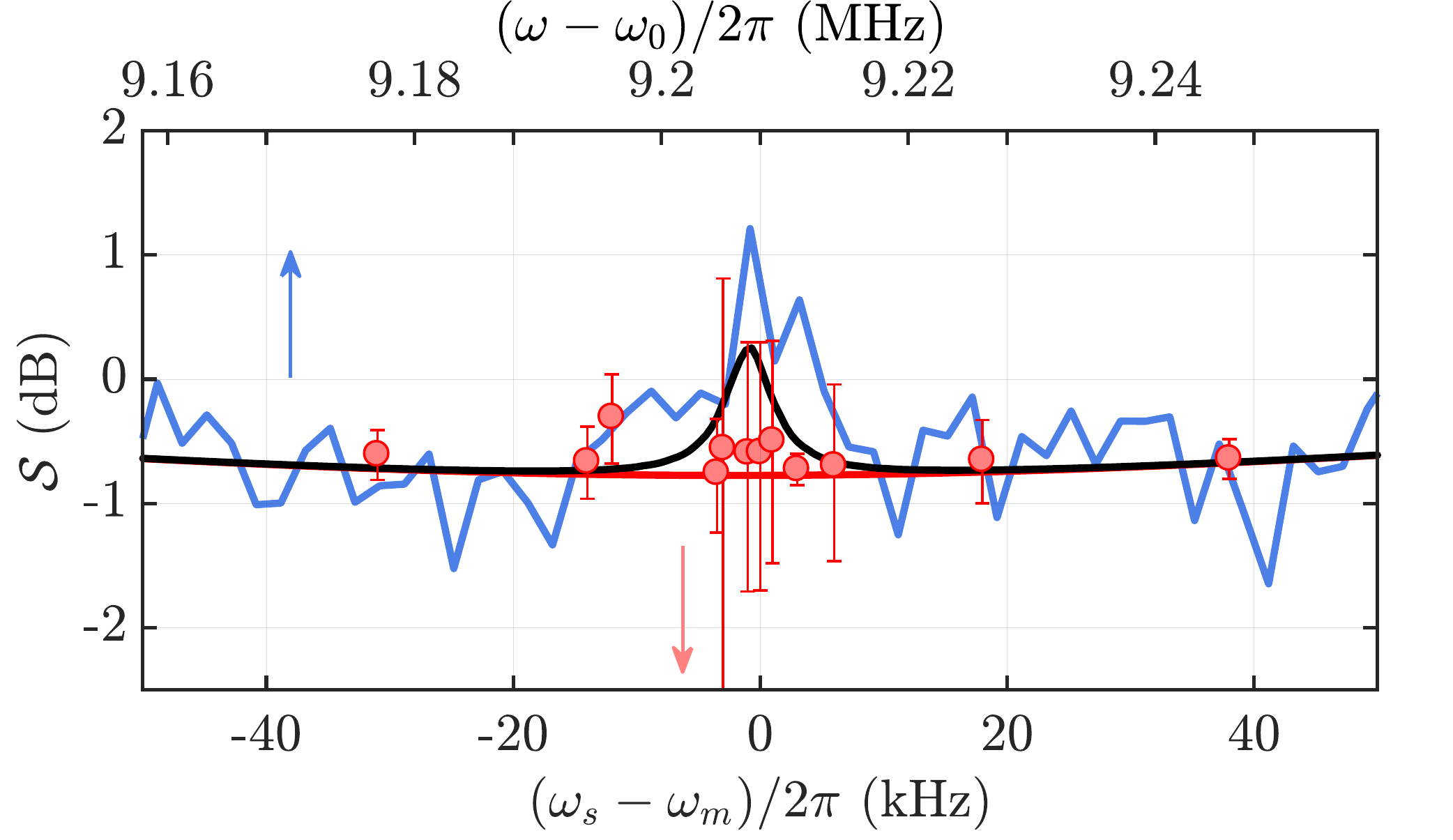}
\caption{\emph{Mapping the correlation matrix.} The solid symbols, with $2\sigma$ error bars, denote the minimum of the complex squeezing spectrum at zero frequency as a function of the bi-LO detuning. This is compared to the theoretical prediction for the smallest eigenvalues of the correlation matrix (red line).  The hidden quantum correlations are given by the data points falling below the lower homodyne envelope (blue trace) and the corresponding theory line (black), around zero detuning. The frequency axis for the homodyne data is given by the upper labels.}
\label{FigCompEnv}
\end{figure}

We now discuss the main result obtained using the bi-chromatic LO (\fref{Fig1}b). The two LO's have frequencies $\pm \omega_s \simeq \pm \omega_m$ symmetrically at both sides of the pump tone. In the general case they have a small detuning from the mechanical sideband, allowing to map the frequency dependence of the correlation matrix, see \eref{eq:bi}.
We create the two LO's digitally and optimize their amplitude and phase to maximize the squeezing around zero frequency. The optimized amplitude ratio of $\alpha_-$ and $\alpha_+$ differs only $0.2$\% from that predicted by the model. In \fref{FigCompSpec} we display the results at different LO detuning values. With the complex detection, we recover squeezing around the zero frequency, which in the lab frame corresponds to the mechanical sidebands of the cavity resonance. The theoretical prediction, using the same parameters as in \fref{FigPonder}, accurately follows the data.  Finally, in \fref{FigCompEnv} we present a summary of the data from \fref{FigCompSpec}. Here, we extract the squeezing at the bottom of the zero-frequency dip. Since there are only a few data points to consider in case of small detuning, the error bars grow larger in this limit. Nonetheless, we can with a good confidence faithfully map the eigenvalues of the correlation matrix within the ``forbidden'' region of $\sim \pm 5$ kHz around the mechanical resonance.

To conclude, we have investigated propagating microwaves to recover quantum correlations that hitherto have remained elusive. Our work also confirms ponderomotive squeezing at a frequency range four orders of magnitude lower than previously demonstrated. The hidden correlations are foreseen to exist and be measurable also in other systems where the output field has to be expressed as a mixture of the quadratures of the field entering the system. In the present system, we expect the reduced noise at a resonant condition to be useful for sensitive force detection.

\begin{acknowledgments} This work was supported by the Academy of Finland (contracts 250280, 308290, 307757, 312296), by the European Research Council (615755-CAVITYQPD), and by the Centre for Quantum Engineering at Aalto University. We acknowledge funding from the European Union's Horizon 2020 research and innovation program under grant agreement No.~732894 (FETPRO HOT). We acknowledge the facilities and technical support of Otaniemi research infrastructure for Micro and Nanotechnologies (OtaNano).
\end{acknowledgments}

\end{document}